\documentclass[prl, twocolumn,superscriptaddress,preprintnumbers,a4paper,amsmath,amssymb,showpacs,floatfix]{revtex4}
\usepackage{graphicx}
\usepackage{color}\color[rgb]{0.000,0.000,0.000} 
\usepackage{amssymb} 
\usepackage{amsmath} 
\usepackage{multirow}

\begin{document}

\newcommand{\ba}{Ba$_{3}$NaRu$_2$O$_9$}
\newcommand{\five}{Ru$^{5+}_2$O$_9$}
\newcommand{\six}{Ru$^{6+}_2$O$_9$}
\newcommand{\te}{$_{t}$}
\newcommand{\ot}{$_{o}$}
\newcommand{\Tc}{T$_{C}$}
\newcommand{\Ts}{T$_{s}$}
\newcommand{\Tn}{$T_\mathrm{N}$}
\newcommand{\MuB}{$\mu_\mathrm{B}$}
\title{Charge order at the frontier between the molecular and solid states in \ba}
\author{Simon A. J. Kimber}
\email[Email of corresponding author:]{kimber@esrf.fr}
\affiliation{European Synchrotron Radiation Facility (ESRF), 6 rue Jules Horowitz, BP 220, 38043  Grenoble Cedex 9, France}
\affiliation{Helmholtz-Zentrum Berlin f\"ur Materialien und Energie (HZB),  Hahn-Meitner Platz 1, 14109, Berlin, Germany}

\author{Mark S. Senn}
\affiliation{School of Chemistry, Joseph Black Building, King's Buildings, West Mains Road, Edinburgh, EH9 3JJ}

\author{Simone Fratini}
\affiliation{Institut N\'eel-CNRS and Universit\'e Joseph Fourier, Bo\^ite Postale 166, F-38042 Grenoble Cedex 9, France}

\author{Hua Wu}
\affiliation{Laboratory for Computational Physical Sciences, Surface Physics Laboratory and Department of Physics, Fudan University, Shanghai 200433, China}

\author{Adrian H. Hill}
\affiliation{European Synchrotron Radiation Facility (ESRF), 6 rue Jules Horowitz, BP 220, 38043  Grenoble Cedex 9, France}

\author{Pascal Manuel}
\affiliation{ISIS Science and Technology Facilities Council, Rutherford Appleton Laboratory, Harwell Science and Innovations Campus, Didcot, OX11 0QX, United Kingdom}

\author{J. Paul Attfield}
\affiliation{School of Chemistry, Joseph Black Building, King's Buildings, West Mains Road, Edinburgh, EH9 3JJ}

\author{Dimitri N. Argyriou}
\affiliation{Helmholtz-Zentrum Berlin f\"ur Materialien und Energie (HZB),  Hahn-Meitner Platz 1, 14109, Berlin, Germany}
\affiliation{European Spallation Source ESS AB, Box 176, 22100, Lund, Sweden}

\author{Paul. F. Henry}
\affiliation{Helmholtz-Zentrum Berlin f\"ur Materialien und Energie (HZB),  Hahn-Meitner Platz 1, 14109, Berlin, Germany}
\affiliation{European Spallation Source ESS AB, Box 176, 22100, Lund, Sweden}

\date{\today}

\begin{abstract}
We show that the valence electrons of \ba, which has a quasi-molecular structure, completely crystallize below 210 K. Using an extended Hubbard model, we show that the charge ordering instability results from long-range Coulomb interactions. However, orbital ordering, metal-metal bonding and formation of a partial spin gap enforce the magnitude of the charge separation. The striped charge order and frustrated $hcp$ lattice of Ru$_{2}$O$_{9}$ dimers lead to competition with a quasi-degenerate charge-melted phase under photo-excitation at low temperature. Our results establish a broad class of simple metal oxides as models for emergent phenomena at the border between the molecular and solid states. \end{abstract}
\maketitle
Strong many body correlations between electrons in solids result in extremely diverse properties including insulating antiferromagnetism, charge order and superconductivity. 
\begin{figure}[tb!]
\begin{center}
\includegraphics[scale=0.3]{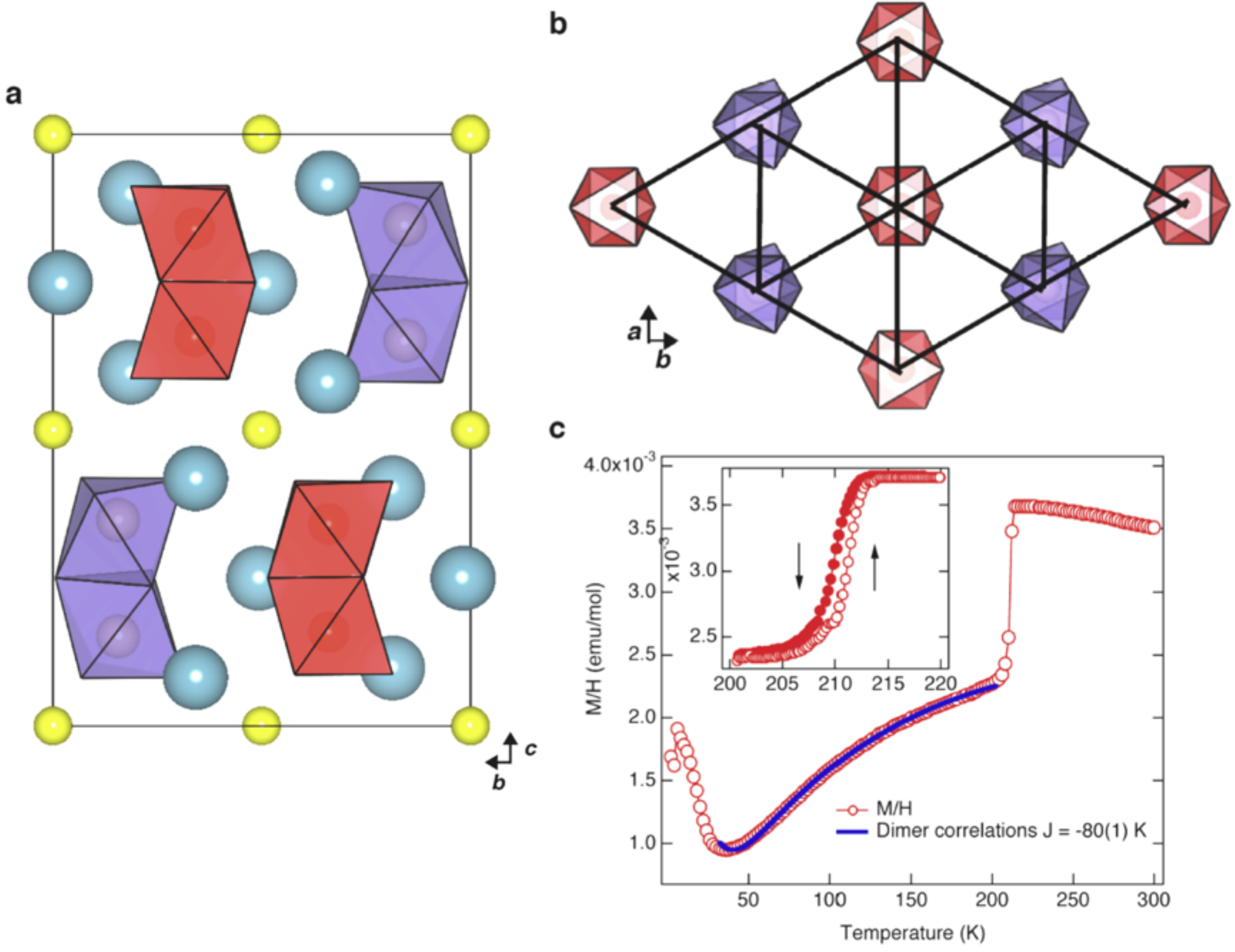}
\caption{(color online) a) CO $P2/c$ structure of \ba \ as determined from synchrotron powder X-ray diffraction projected down  [100], Sodium atoms are yellow spheres, Barium atoms are  green. The symmetry inequivalent Ru$_{2}$O$_{9}$ dimers are shown in red (\five) and in blue (\six). (b) $P2/c$ structure projected down [001] showing the frustrated triangular layers. (c) Results of magnetic susceptibility measurements for \ba \ in a 1000 Oe field, showing fit to Ru$^{6+}$  dimer correlations below 210 K. The inset (same units) shows the hysteresis on warming/cooling.}
\label{Fig1}
\end{center}
\end{figure}
This is especially true in organic based materials like fullerides or the $\kappa$-$(BEDT$-$TTF)_{2}$X or (Pd[dmit]$_{2}$)$_{2}$ salts, due to the narrow electronic bandwidth resulting from the overlap of molecular orbitals \cite{coronado,kurmoo,powell}. However, the contacts between molecules are weak, and organic bonding allows low-energy rotational and conformational degrees of freedom. Poorly crystalline or complex, low symmetry structures \cite{ravy,rouziere,nakao} are thus common. This problem can be compounded by extreme air sensitivity, the invisibility of hydrogen to X-ray diffraction, and the incoherent background which it generates in neutron scattering experiments. Consequently, nearly all theoretical models of these materials rely on approximations that treat each molecular unit as effective sites on high symmetry lattices \cite{powell,scriven}. These models predict that Coulomb interactions drive electron localisation, and that geometrical frustration is responsible for the observed competition between ground states \cite{kino}. However, the material-specific interplay between electronic correlations, local-degrees of freedom, and the lattice remains a significant challenge \cite{kandpal}. \\
In this Letter, we show that simple metal oxides can provide new insight into the electronic properties which emerge between the molecular and solid states. Our model compound is \ba, which has the 6H hexagonal perovskite structure (Figs. 1a and 1b) containing isolated Ru$^{5.5+}_{2}$O$_{9}$ dimers of face-sharing octahedra which form triangular layers \cite{stitzer}. Previous reports have indicated poorly-metallic conduction around room temperature and a metal-insulator transition \cite{quarez} \ associated with a structural distortion below 210 K.
We synthesised single crystal and powder samples of \ba \  which were characterised by magnetisation and high resolution synchrotron X-ray and time-of-flight neutron diffraction measurements. We confirmed that the room temperature structure has hexagonal $P6_{3}/mmc$ symmetry in which all Ru sites are equivalent.  The magnetic susceptibility was observed to approximately halve on cooling through a first-order transition at T$_{co}$ = 210 K, consistent with the opening of a charge or spin gap (Fig. 1c). This is accompanied by the formation of an apparently $C$-centred, orthorhombic 2.\textit{\textbf{a}} x 2.$\surd$3\textit{\textbf{a}} x \textit{\textbf{c}} superstructure \cite{stitzer}, however our high resolution X-ray synchrotron powder diffraction measurements, performed using ID31 at the ESRF, detected a previously unreported monoclinic distortion (Fig. 2a) and further weak reflections that showed that the true space group is $P2/c$. The structural model was refined using X-ray data from a twinned single-crystal, and full details are given in the supplemental material.\\
The key feature of our low temperature $P2/c$ structure is a splitting of the Ru$^{5.5+}_{2}$O$_{9}$ dimers into two symmetry inequivalent units, which suggests charge ordering (CO) (Fig. 1b). Motivated by experiments on true molecular materials, we investigated the driving force for this transition using the following extended Hubbard model, with the dimers as effective sites.
\begin{equation*}
\hat{H}=-t\sum_{<ij>\sigma}(c_{i\sigma}^{\dagger}c_{j\sigma}+h.c)+U_{eff}\sum_{i}n_{i\uparrow}n_{i\downarrow}+\sum_{<ij>}V_{ij}n_{i}n_{J}
 \end{equation*}			
Here $t$ \ is the interdimer hopping integral, $U_{eff}$ is the on-site (dimer) Coulomb repulsion and $V$ \ is the interdimer Coulomb repulsion. In contrast to Mott insulating organic materials such as the $\kappa$-$(BEDT$-$TTF)_{2}$X  family, where $V$ \ is commonly neglected \cite{kandpal}, the experimentally determined stripe order could only be reproduced when $V > U_{eff}$. This implies that the formation of molecular orbitals results in a reduced $U_{eff}$ \cite{scriven}, and that the inter-site electron-electron interactions play a fundamental role in the ordering mechanism. However, without breaking the three-fold lattice symmetry, a striped groundstate is still highly degenerate. 
\begin{figure}[tb!]
\begin{center}
\includegraphics[scale=0.155]{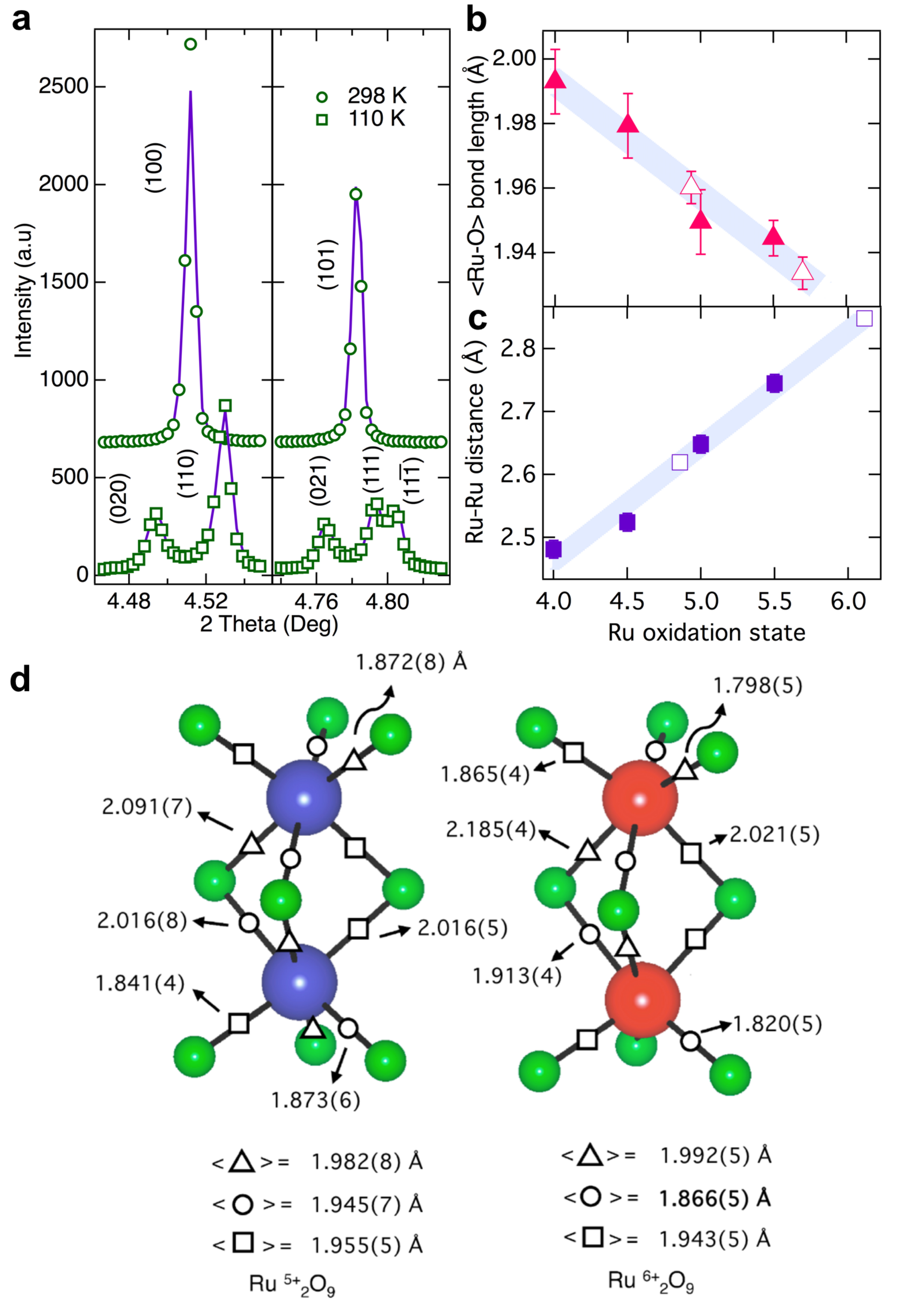}
\caption{(color online) (a) Splitting and peak broadening observed in \ba \ by synchrotron powder X-ray diffraction upon cooling into the $P2/c$ CO phase. (b) The Ru-Ru distance has a linear dependence on oxidation state in the Ba$_{3}$ARu$_{2}$O$_{9}$ \ family of compounds with A = Na$^{+}$, Ca$^{2+}$, Nd$^{3+}$ and Ce$^{4+}$, which have similar ionic radii. The formal ruthenium oxidation states are 5.5$^{+}$, 5$^{+}$, 4.5$^{+}$ and 4$^{+}$, open symbols are from the $P2/c$ \ phase determined in this work. The shaded region shows the standard deviation of the calibration fit. (c) A similar linear dependence upon formal oxidation state is found for the average Ru-O bond distance in the  Ba$_{3}$ARu$_{2}$O$_{9}$ \ materials. The symbols are as above. (d) Coordination environment in the two inequivalent Ru$_{2}$O$_{9}$ dimers. The Ru$^{6+}$ octahedra are compressed implying $t_{2g}^{2}$ orbital ordering.}
\label{Fig1}
\end{center}
\end{figure}
We discovered that the full three-dimensional $P2/c$ structure could be rationalised by considering the tetrahedral linkages between the $hcp$ network of dimers. Each tetrahedron of dimers contains two \five \ and two \six \ units (S1). Hence, so-called 'ice rules', which minimise the electrostatic energy of the molecular building blocks \cite{anderson,udagawa}, act as directors for the fluctuating stripe order stabilised by the reduced $U_{eff}$. On the macroscopic scale, the twinning in single crystal samples has exactly the degeneracy predicted by this picture, as six orientational stripe domains of equal population are found. Our observations thus establish the Ru$_{2}$O$_{9}$ dimers as the physically meaningful units in \ba. This analysis makes no prediction of the size of the charge separation between the sites. However, our precise refinement of $P2/c$ structure allowed us to examine structural parameters on the single-ion and molecular levels, and compare them to other members of the Ru$_{2}$O$_{9}$ family of known oxidation states. The valence of each individual Ru cation is reflected \cite{stitzer} \ by a linearly varying $<$Ru-O$>$ distance (Fig. 2b), and the refined values for the $P2/c$ structure are in agreement with 'integer' separation into 5$^{+}$ and 6$^{+}$ oxidation states. Furthermore, detailed examination of the coordination environment (Fig. 2d) of the two sites provides evidence for ordering of orbital degrees of freedom. In order to separate out the effect of the intrinsic trigonal site distortion from electronic effects, we calculated the average length of perpendicular bond pairs. For the Ru$^{5+}$ sites, which are orbitally non-degenerate with a $t_{2g}^{3}$ configuration, these are extremely regular, and lie in the range 1.944(5)-1.9815(8) \AA. In contrast, the $t_{2g}^{2}$  Ru$^{6+}$ sites have four long bonds with average lengths in the range 1.943(5)-1.991(5) \AA \ and two short bonds of average length 1.866(5) \AA. An orbital ordering distortion therefore lifts the degeneracy of the doubly occupied $t_{2g}$ \ orbitals. In addition to these single-ion indicators of integer charge order, the interdimer Ru-Ru distances also vary linearly with oxidation state (Fig. 2c). This implies a well defined 'bond order' determined by the occupation of molecular orbital states and allows a precise estimate of the charge separation into 4.86(16)+ and 6.11(16)+ states. The Ru-Ru bond in the \five \ dimers is as short as that found in Ru metal (2.62 \AA). Metallic bonding would make these sites non-magnetic, which is in accordance with the magnetic susceptibility below 210 K, which could be modelled assuming only \textbf{S}=1 correlations from the \six \ dimers (Fig. 1c).
Finally, we used the precise $P2/c$ coordinates to perform LDA+U calculations with U = 3 eV, which confirm that complete disproportionation into \five \ and \six \ pairs is found (Fig. 3a). This is stabilised by the \textbf{S} = 0 dimer states, as antiferromagnetic coupling gives by far the lowest energy groundstate. A small gap at the Fermi energy is found in accordance with Ref. \cite{quarez}. These calculations confirm that the simplest molecular constraint, the desire to form closed shell configurations, stabilises the complete charge segregation found below 210 K . We emphasise that our experimental and theoretical quantification of the degree of charge localisation in \ba \ would have been effectively impossible for organic materials. \\
The integer charge separation motivates comparison with frustrated Ising spin models on the $hcp$ lattice, which have the same groundstate configuration \cite{auerbach}. Geometrical frustration of  electronic degrees of freedom in molecular solids is additionally predicted to result in competition between charge ordered and disordered states \cite{udagawa,merino}. Although the $P2/c$ phase of \ba \  has large ($\sim$0.1 \AA) atomic displacements, we discovered that the charge order is unusually sensitive to external perturbations. A comparison between the lattice parameters extracted from X-ray (ID31 at ESRF) and neutron (WISH at ISIS) diffraction experiments is shown in Fig. 3b. Below $\sim$40 K these deviate strongly from each other. This was investigated by cooling a sample to 10 K before continuously measuring diffraction profiles with 31 keV X-rays. We found that the $P2/c$ charge ordered structure transforms continuously to a higher symmetry $C2/c$ structure in which only one Ru site is present. This phase was found to be stable for periods of several hours at 10 K, during and after irradiation, and was not observed by neutron powder diffraction down to 1.6 K, which also showed no evidence for long-range magnetic order. Band structure calculations predict metallic conduction in this phase.\\
\begin{figure}[tb!]
\begin{center}
\includegraphics[scale=0.32]{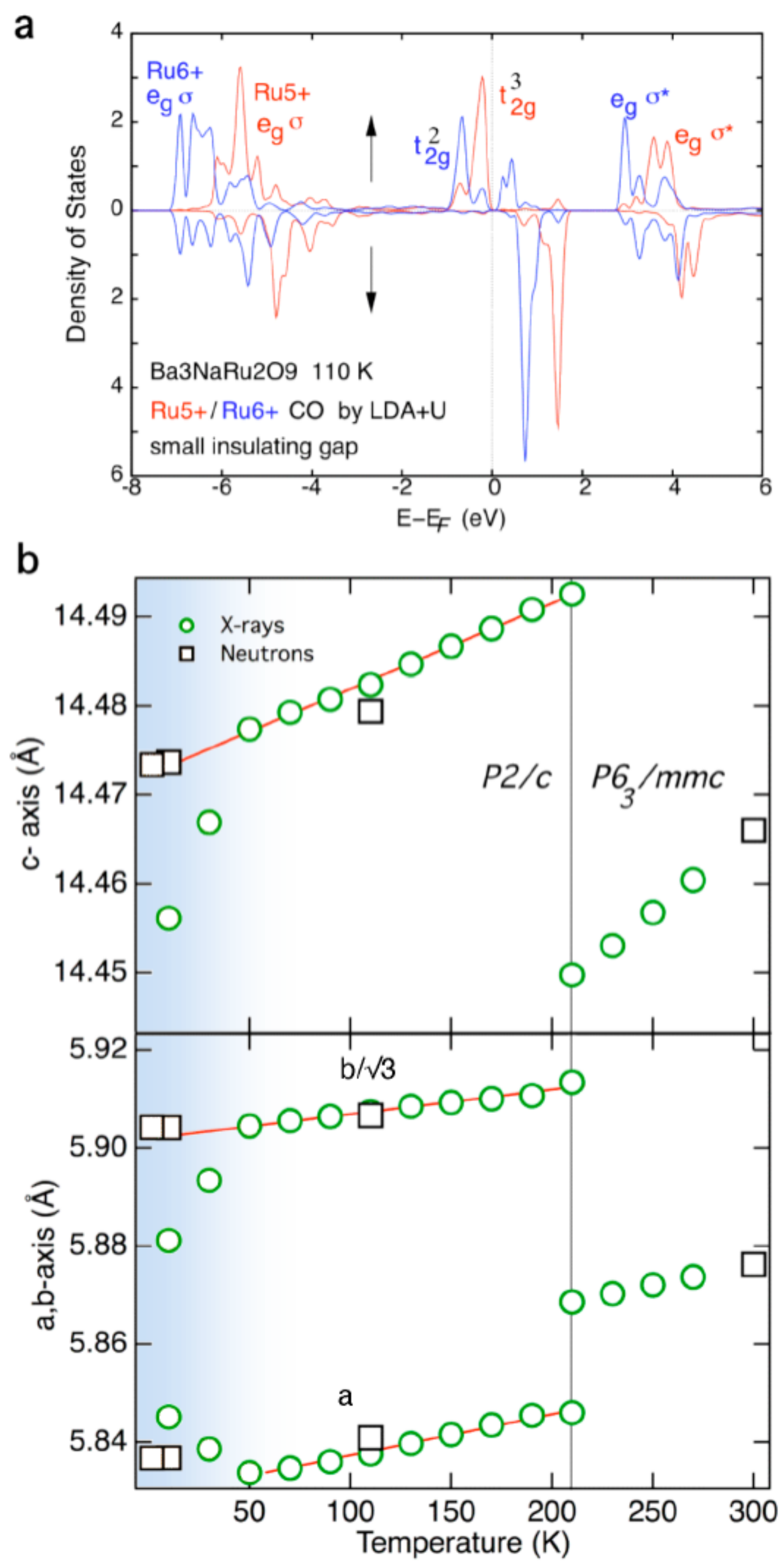}
\caption{(color online) (a) Site specific ruthenium density of states determined by LDA+U calculations for the charge ordered $P2/c$ phase of \ba, a small insulating gap is stabilised, in accordance with Ref. (12). (b) Temperature dependence of lattice parameters determined by Rietveld refinement of powder synchrotron X-ray and neutron powder diffraction. Lines show fits used to extrapolate T = 0 data points used in Fig. 4}
\label{Fig1}
\end{center}
\end{figure}
The synchrotron X-ray data collected during the melting process show that local and lattice degrees of freedom are inextricably linked. As shown in Fig. 4a, the average Ru$^{5+}$/Ru$^{6+}$ charge order, quantified by the difference between the Ru-Ru distances in the two dimers, decreases rapidly with an exponential time constant of 75(8) s. After 120-180 s, this charge order and the corresponding ($h+k$) = $odd$ superstructure reflections are completely suppressed showing that the average structure has transformed to $C2/c$ symmetry. The very high resolution of the synchrotron diffraction data makes a transition to an intermediate charge glass of localised but spatially disordered charges unlikely\ \cite{ishibashi}, as the microscopic lattice strain, which measures local variations in lattice parameters, also rapidly vanishes. The lattice distortion which breaks frustration, parameterised by the \textit{\textbf{b}}/\textit{\textbf{a}} ratio, is coupled to the local order, and relaxes on a similar timescale (88(3) s). \\
A possible explanation for the CO melting is the dimensional reduction into stripes \cite{hotta} \ patterned by the ice-rule constraints. X-rays impinging on the sample generate a low concentration of holes by the photoelectric effect. Based on our minimal electrostatic model, these may freely delocalise along the stripe direction, provided that $t \geqslant V-U_{eff}/4$, melting the charge order. That this Ôcharge meltedÕ phase exists in delicate balance with the charge order, is shown by measurements on warming, as the  $P2/c$ phase is recovered on warming above 40 K with full recovery of the lattice micro and macrostrains. Future work will be needed to determine the factors which set this temperature scale. However, we note that $\chi(T)$ also reproducibly shows a glassy transition \ \cite{stitzer} at similar temperatures (Fig. 1c), which might indicate an intrinsic instability of the CO, even in the absence of irradiation. \\
\begin{figure}[tb!]
\begin{center}
\includegraphics[scale=0.24]{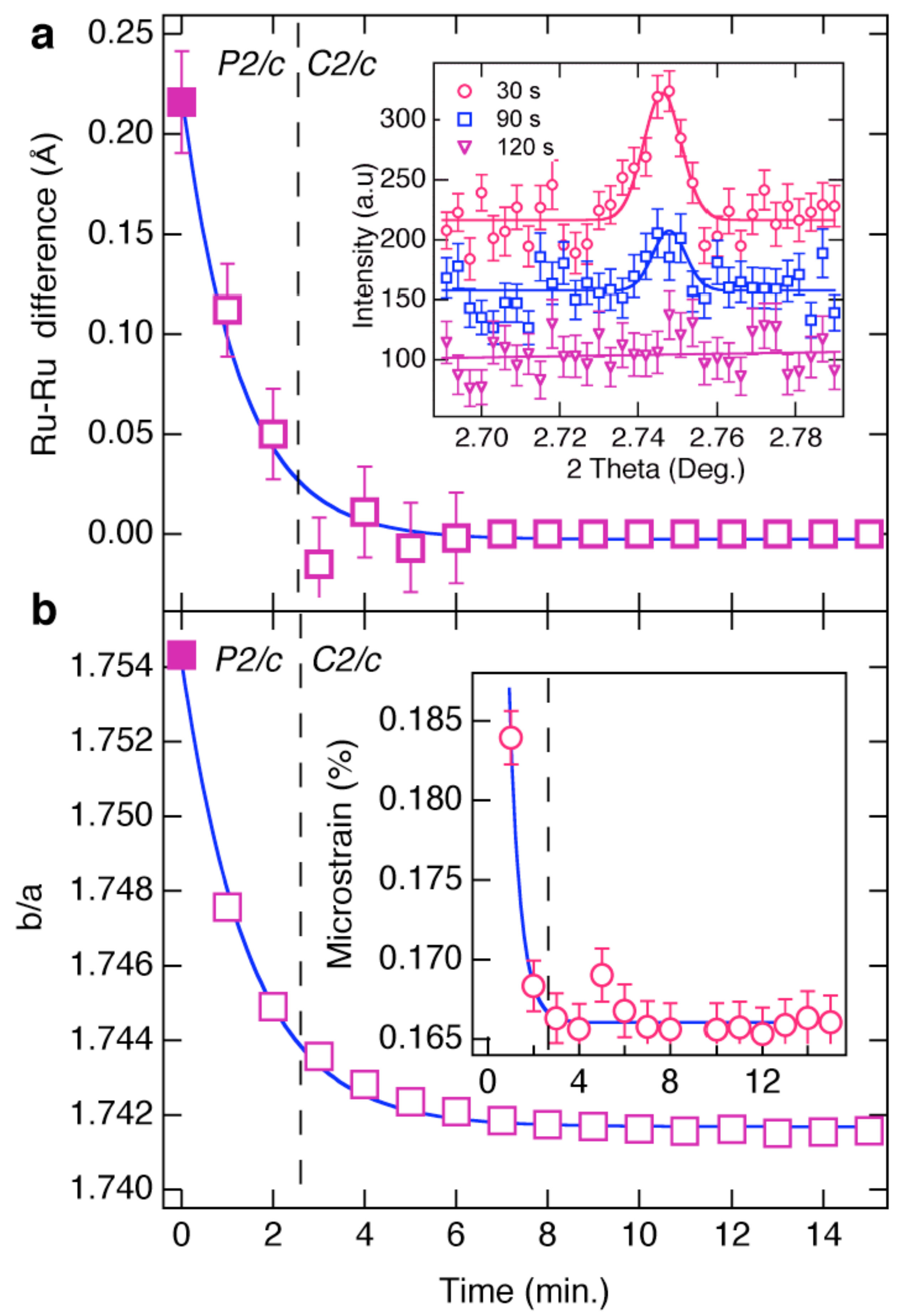}
\caption{(color online) (a) Difference between the refined Ru-Ru distance for the two dimer types in \ba as a function of irradiation time at 10 K. The ($h+k$)=$odd$ superstructure reflections (inset) also disappear, showing that the CO melts. Results of fitting exponential relaxation functions are shown; (b) Time dependence of the lattice macrostrain (\textit{\textbf{b}/\textbf{a}}) ratio. The inset shows the immediate decrease in the lattice microstrain, determined from the tan$\theta$ dependent contribution to the Lorentzian peak shape.}
\label{Fig1}
\end{center}
\end{figure}
Unlike other charge-ordered metal oxides, \ba \ shows both fully crystallized and liquid-like valence electron states at low temperatures. However, the classic examples of oxide charge order are all found in structure types which have infinite lattices of corner, edge or face sharing coordination polyhedra, like perovskites or spinels. The mechanism for charge order is somewhat controversial in these materials, as only fractional charge separation is found \cite{attfield}, with strong evidence for Fermi surface nesting \cite{milward}.  In contrast,  our results shows that long-range electrostatic interactions are responsible for charge ordering in \ba. Additionally, the comparatively simple crystal structures highlight the role played by single-ion and molecular degrees of freedom. Our results thus lend weight to a recent theoretical investigation of $\kappa$-ET$_{2}$Cu$_{2}$(CN)$_{3}$, which showed that intra-molecular degrees of freedom must be important in this related class of materials \cite{hotta2}.\\
In summary, our results show that a combination of single-ion and molecular degrees of freedom help to stabilise charge order in \ba. Our results are reminiscent of the true (organic) molecular solids and provide a model system for studying frustrated strongly correlated electrons. We note the presence of a vast library of unexplored related compounds \cite{mitchell}, and predict the future discovery of strongly fluctuating spin-liquid or superconducting groundstates.\\
We thank the ESRF and ISIS for access to their instruments and the HZB, EPSRC and Leverhulme trust for support. We thank S. Parsons for assistance with the single crystal data collection, and C.D. Ling, R. Valent\'i, V. Honkim\"aki and T. Chatterji for useful discussions. 

\begin{figure*}[tb!]
\begin{center}
\includegraphics[scale=1]{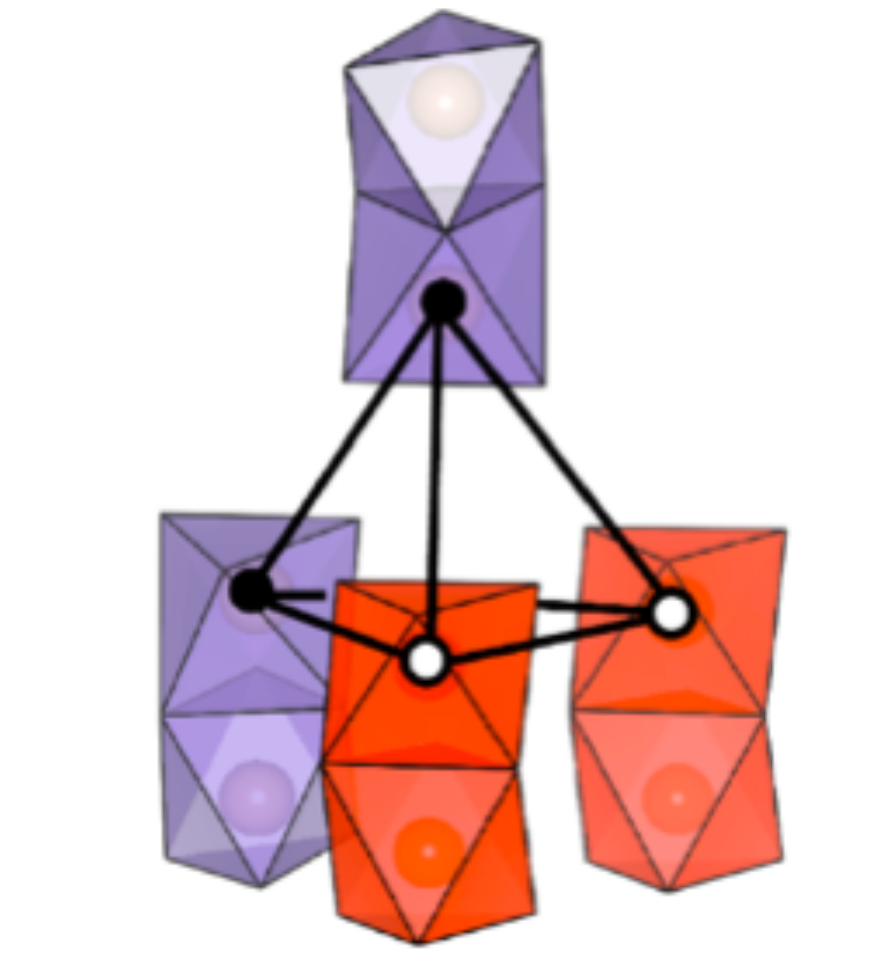}
\caption{(color online) \textbf{S1} \ The $hcp$ lattice of Ru$_{2}$O$_{9}$ dimers in Ba$_{3}$NaRu$_{2}$O$_{9}$ has tetrahedral linkages between dimers in the stacking (\textit{\textbf{c}}) direction. In the charge ordered phase, two of each charge state are found in each tetrahedral unit, which minimises the electrostatic energy of the charge ordered state.}
\label{S1}
\end{center}
\end{figure*}
\begin{table*}
\caption{\label{tab:table2}Details of single crystal X-ray data collection and refinement of \ba \ at 110 K}
\begin{ruledtabular}
\begin{tabular}{cc}
 $$ &$$\\
\hline
Formula & \ba \\
Cell setting, Space group                               &   Monoclinic, $P 1 2/c 1$\\
Temperature(K)                                          &   110 K (2)\\
a,b,c (\AA)                                             &   5.84001(2) , 10.22197(4) , 14.48497(6)\\
$\beta$                                                 &   90.2627(3)\\
Volume (\AA$^{3}$)                                      &   864.69(2)\\
D$_{c}$ (mg.m$^{-3}$)                                   &   6.00\\
Radiation type                                          &   Mo-K$\alpha$\\
No. reflections for cell parameters                     &   725\\
$\mu$ (mm$^{-1}$)                                       &   16.948\\
Crystal form, colour                                    &   hexagonal plate, black\\
Crystal size (mm)                                       &   $0.13 \times 0.45 \times 0.45$\\
Diffractometer                                          &   Bruker Apex2\\
Data collection method                                  &   $\omega+\varphi$\\
Absorption correction                                   &   Spherical\\
$T_\mathrm{min}$                                        &   0.0616\\
$T_\mathrm{max}$                                        &   0.1174\\
No. of measured, independent and observed reflections   &   11855, 2549\\
$\Theta_\mathrm{max}$($^{\circ}$)                       &   30.5293\\
R$_{int}$                                               &   0.067\\
Range of h,k,l                                          &   $\begin{pmatrix} -8 & h & 8 \\ 0 & k & 14 \\ 0 & l &20 \end{pmatrix}$\\
Refined \ on                                            &   F\\
R, wR(F), S                                             &   0.028, 0.032, 1.01\\
Cutoff: I $>$                                           &   3.00$\sigma$(I)\\
No. of reflections                                      &   2651\\
No. of parameters                                       &   80\\
\multirow{4}{*}{Weighting scheme}                       &   $w=w' \times \left[1 - \left(\Delta F_\mathrm{obs} / 6 \times \Delta F_\mathrm{est}\right)^{2}\right]^{2}$,\\
                                                        &   $w'=[P_{0}T_{0}'(x)+P_{1}T_{1}'(x)+\ldots P_{n-1}T_{n-1}'(x)]^{-1}$,\\
                                                        &   where $P_{i}$ are the coefficients of a\\
                                                        &   Chebychev series in $t_{i}(x)$, and $x = F_\mathrm{calc} /F_\mathrm{calcma}$\\
($\Delta / \alpha$)max                                  &   0.0010\\
$\Delta \rho_\mathrm{max}$, $\Delta \rho_\mathrm{min}$ (e \AA$^{-3}$)    &   1.97, -1.36\\
 \end{tabular}
\end{ruledtabular}
\end{table*}
\begin{table*}
\caption{\label{tab:table2}Twin laws and refined twin fractions}
\begin{ruledtabular}
\begin{tabular}{cc}
 $Twin \  law$ &$Twin \ fraction$\\
\hline
$\begin{pmatrix} 1 & 0 & 0 \\ 0 & 1 & 0 \\ 0 & 0 &1 \end{pmatrix}$&0.1656(15)\\
$\begin{pmatrix} -0.5 & -0.5 & 0 \\ 1.5 & -0.5 & 0 \\ 0 & 0 &1 \end{pmatrix}$&0.165(3)\\
$\begin{pmatrix} -0.5 & 0.5 & 0 \\ -1.5 & -0.5 & 0 \\ 0 & 0 &1 \end{pmatrix}$&0.168(3)\\
$\begin{pmatrix} 1 & 0 & 0 \\ 0 & 1 & 0 \\ 0 & 0 &-1 \end{pmatrix}$&0.1674(15)\\
$\begin{pmatrix} -0.5 & -0.5 & 0 \\ 1.5 & -0.5 & 0 \\ 0 & 0 &-1 \end{pmatrix}$&0.169(3)\\
$\begin{pmatrix} -0.5 & 0.5 & 0 \\ -1.5 & -0.5 & 0 \\ 0 & 0 &-1 \end{pmatrix}$&0.169(3)\\
   \end{tabular}
\end{ruledtabular}
\end{table*}
\begin{table*}
\caption{\label{tab:table2}Refined atomic coordinates and displacement parameters for \ba \ at 110 K from single crystal X-ray diffraction. Anisotropic displacement parameters were only refined for the barium atoms.}
\begin{ruledtabular}
\begin{tabular}{ccccc}
 $Atom$ &$x$ &$y$ &$z$ &$U_{iso/equiv}$\\
\hline
  Na(1)& 0 & 0 & 0 & 0.0044(5)\\
  Na(2)& 0.5 & 0.5 & 0 & 0.0052(5)\\
  Ba(1)& 0 & 0.00689(4) & 0.25 & 0.0051\\
  Ba(2)& 0.5 &0.48195(5) &0.25 & 0.0052\\
  Ba(3)&0.50537(14) & 0.15771(3) & 0.91817(2) & 0.0053\\
  Ba(4)&0.00413(14) & 0.68674(4) & 0.90174(2) & 0.0057\\
  Ru(1)&0.49931(19)&0.16087(3)&0.15941(2)&0.00395(9)\\
  Ru(2)&0.00241(9)&0.65572(4)&0.15151(3)&0.00426(9)\\
  O(1)&0.7317(6)&0.2404(4)&0.2461(6)&0.0054(8)\\
O(2)&0.2065(7)&0.7367(4)&0.2361(3)&0.0040(8)\\
O(3)&0&0.4843(6)&0.75&0.0095(12)\\
O(4)&0.5&-0.0109(6)&0.75&0.0096(12)\\
O(5)&0.2694(14)&0.0793(9)&0.4087(5)&0.0128(19)\\
O(6)&0.8090(9)&0.5704(5)&0.4236(3)&0.0083(9)\\
O(7)&0.0074(15)&0.1905(4)&0.9177(3)&0.0045(7)\\
O(8)&0.4920(16)&0.6846(4)&0.9122(3)&0.0040(7)\\
O(9)&0.7365(10)&0.0857(7)&0.4108(4)&0.0025(14)\\
O(10)&0.2827(8)&0.6028(5)&0.3892(3)&0.0080(9)\\
   \end{tabular}
\end{ruledtabular}
\end{table*}
\end{document}